\documentclass[aps,prb,reprint,longbibliography,superscriptaddress]{revtex4-1}
\usepackage{mathrsfs}
\usepackage{amsmath,gensymb}
\usepackage{amsfonts}
\usepackage{amssymb}
\usepackage{amsthm}
\usepackage{graphicx}
\usepackage{natbib}
\usepackage{color}
\usepackage{hyperref}
\usepackage{bm}
\usepackage[caption=false]{subfig}
\usepackage{verbatim}

\newcommand{\sqket}[1]{\left| {#1} \right\rangle_{\mathrm{sq}}}
\newcommand{\subket}[1]{\left| {#1} \right\rangle_{\mathrm{sub}}}
\newcommand{\sqbra}[1]{\left\langle {#1} \right|_{\mathrm{sq}}}
\newcommand{\subbra}[1]{\left\langle {#1} \right|_{\mathrm{sub}}}

\begin{document}
\title{Antiferromagnetic magnons as highly squeezed Fock states underlying quantum correlations}

\author{Akashdeep Kamra}
\email{akashdeep.kamra@ntnu.no}
\affiliation{Center for Quantum Spintronics, Department of Physics, Norwegian University of Science and Technology, Trondheim, Norway}

\author{Even Thingstad}
\affiliation{Center for Quantum Spintronics, Department of Physics, Norwegian University of Science and Technology, Trondheim, Norway}

\author{Gianluca Rastelli}
\affiliation{Department of Physics, University of Konstanz, Konstanz, Germany}
\affiliation{Zukunftskolleg, University of Konstanz, Konstanz, Germany}

\author{Rembert A. Duine}
\affiliation{Utrecht University, Utrecht, The Netherlands}
\affiliation{Department of Applied Physics, Eindhoven University of Technology, Eindhoven, The Netherlands}
\affiliation{Center for Quantum Spintronics, Department of Physics, Norwegian University of Science and Technology, Trondheim, Norway}

\author{Arne Brataas}
\affiliation{Center for Quantum Spintronics, Department of Physics, Norwegian University of Science and Technology, Trondheim, Norway}

\author{Wolfgang Belzig}
\affiliation{Department of Physics, University of Konstanz, Konstanz, Germany}

\author{Asle Sudb{\o}}
\affiliation{Center for Quantum Spintronics, Department of Physics, Norwegian University of Science and Technology, Trondheim, Norway}

\begin{abstract}
Employing the concept of two-mode squeezed states from quantum optics, we demonstrate a revealing physical picture for the antiferromagnetic ground state and excitations. Superimposed on a N{\'e}el ordered configuration, a spin-flip restricted to one of the sublattices is called a sublattice-magnon. We show that an antiferromagnetic spin-up magnon is comprised by a quantum superposition of states with $n+1$ spin-up and $n$ spin-down sublattice-magnons, and is thus an enormous excitation despite its unit net spin. Consequently, its large sublattice-spin can amplify its coupling to other excitations. Employing von Neumann entropy as a measure, we show that the antiferromagnetic eigenmodes manifest a high degree of entanglement between the two sublattices, thereby establishing antiferromagnets as reservoirs for strong quantum correlations. Based on these insights, we outline strategies for exploiting the strong quantum character of antiferromagetic (squeezed-)magnons and give an intuitive explanation for recent experimental and theoretical findings in antiferromagnetic magnon spintronics.
\end{abstract}

\maketitle

\section{Introduction}

As per the Heisenberg uncertainty principle, the quantum fluctuations of two non-commuting observables cannot simultaneously be reduced to zero. However, it is possible to generate a state with the quantum noise in one observable reduced below its ground state limit at the expense of enhanced fluctuations in the other observable~\cite{Gerry2004,Schnabel2017}. Considering a single mode or frequency of light, such states, generally called squeezed vacuum~\cite{Gerry2004,Schnabel2017}, have proven instrumental in the detection of gravitational waves~\cite{LIGO2016} with a sensitivity beyond the quantum ground state limit~\cite{LIGO2011,LIGO2013,Grote2013}. Furthermore, squeezed vacuum states have applications in quantum information~\cite{Ou1992,Ralph1999,Milburn1999,Furrer2012,Eddins2018} since they exhibit quantum correlations and entanglement. These are best represented and exploited via the two-mode squeezed vacuum states, where the two participating modes are entangled and correlated~\cite{Gerry2004}. The widely studied~\cite{Gerry2004,Schnabel2017} single- and two-mode squeezed vacuums may be considered a special case, corresponding to zero photon number(s), of a wider class - squeezed Fock states~\cite{Kral1990,Nieto1997}. While investigated theoretically, the latter have been largely forgotten, probably owing to the experimental challenge of generating them. The squeezing concept applies to bosonic modes in general, and squeezed states of magnons~\cite{Zhao2004,Zhao2006,Bossini2016,Bossini2019} and phonons~\cite{Johnson2009} have also been achieved experimentally. 

The concept of squeezed Fock states~\cite{Kral1990,Nieto1997} has proven valuable in understanding the spin excitations of ordered magnets~\cite{Kamra2016A,Kamra2017A}.  Squeezed-magnons have been shown to be the eigen-excitations of a ferromagnet~\cite{Kamra2016A,Kamra2016B}. A squeezed-magnon is comprised by a coherent superposition of the different odd number states of the spin-1 magnon~\cite{Kamra2016A,Kamra2017A}~\footnote{The ``spin-1'' magnon is a quasiparticle that carries a spin of $\hbar$ along the z-direction~\cite{Kittel1963}. It is not an actual $S=1$ bosonic particle.}. This bestows it a noninteger average spin larger than 1. The relatively weak spin-nonconserving interactions, such as dipolar fields or crystalline anisotropy, underlie the magnon-squeezing in ferromagnets. These spin-nonconserving interactions were further found to result in two-sublattice magnets hosting excitations with spin varying continuously between positive and negative values~\cite{Kamra2017A}. In contrast, exchange interaction in a two-sublattice magnet leads to a strong squeezing effect, which does not affect the excitation spin and forms a main subject of the present article. Being eigen-excitations, squeezed-magnons are qualitatively distinct in certain ways from the squeezed states of light discussed above, which are non-equilibrium states generated via an external drive. At the same time, the two kinds of states share several similar features on account of their wavefunctions being mathematically related. To emphasize this difference, we employ the terminology that ``squeezed state of a boson'' refers to a non-equilibrium state, while a ``squeezed-boson'' is an eigenmode~\footnote{Within the adopted terminology convention, if one were to generate a non-equilibrium squeezed state of spin excitations in an anisotropic ferromagnet, it would be called ``squeezed state of squeezed-magnons''.}. 

Instigated by recent experimental breakthroughs~\cite{Saitoh2006,He2010,Zhang2016,Wadley2016,Kosub2017,Lebrun2018}, interest in antiferromagnets (AFMs) for practical applications has been invigorated~\cite{Gomonay2014,Jungwirth2016,Gomonay2018,Baltz2018,Libor2018}. Due to the well-known strong quantum fluctuations in AFMs, they have also been the primary workhorse of the quantum magnetism community~\cite{Sachdev2001}. The N{\'e}el ordered configuration, which is consistent with most of the experiments, is not the true quantum ground state of an AFM. Furthermore, quantum fluctuations destroy any order in a one-dimensional isotropic AFM. These and related general ideas applied to AFMs bearing geometrically frustrated interactions underlie quantum spin liquids~\cite{Castelnovo2008,Balents2010,Savary2016}, which are devoid of order in the ground state and host exotic, topologically non-trivial excitations embodying massive entanglement. 

\begin{figure}[tb]
	\begin{center}
		\subfloat[]{\includegraphics[width=60mm]{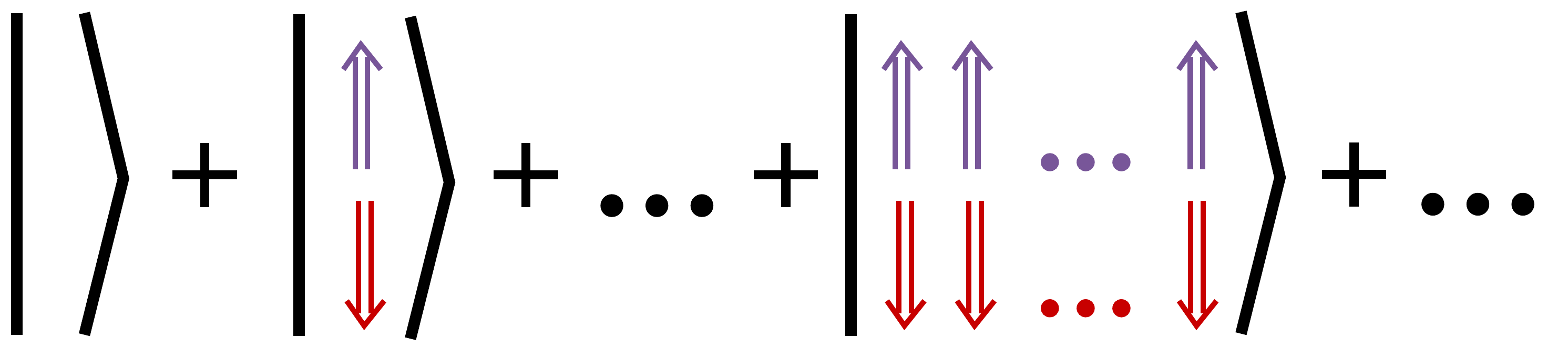}}\\
		\subfloat[]{\includegraphics[width=60mm]{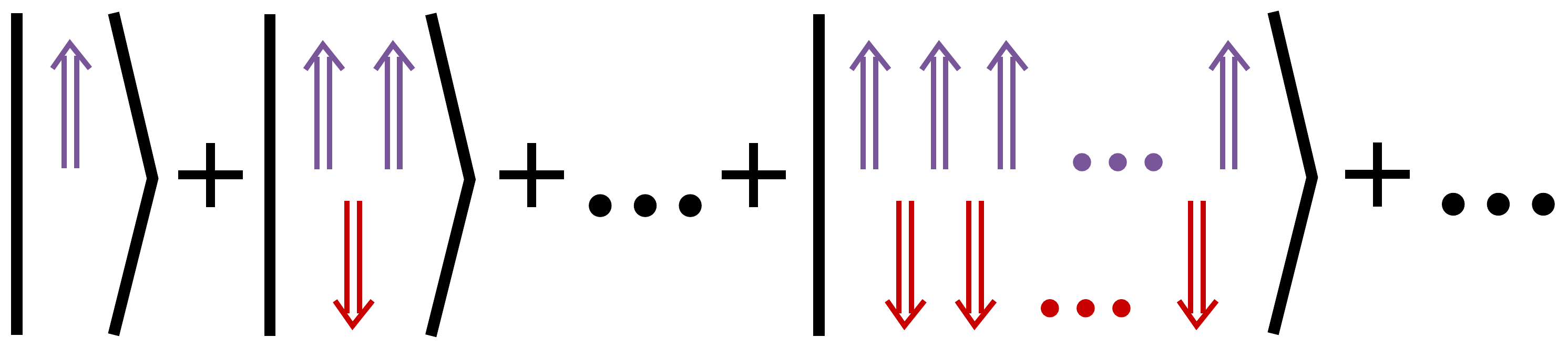}}\\
		\subfloat[]{\includegraphics[width=70mm]{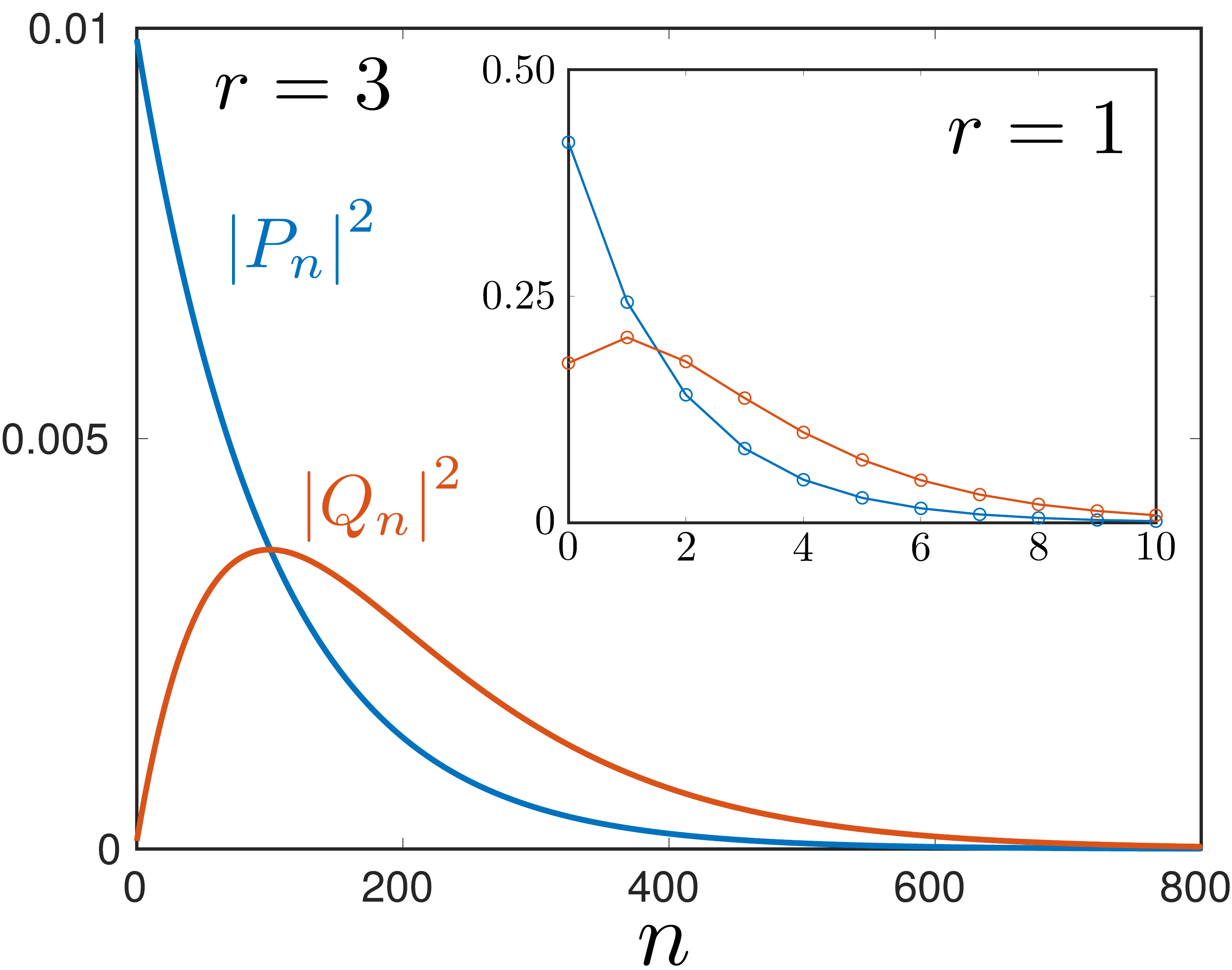}}
		\caption{Schematic depiction of spatially uniform antiferromagnetic (a) vacuum and (b) spin-up eigenmodes. (a) The vacuum mode, represented as $\sqket{0} = \sum_n P_n \subket{n,n}$, is a superposition over states with equal number of spin-up and -down sublattice-magnons. (b) The spin-up squeezed-magnon, represented as $\sqket{\uparrow} = \sum_n Q_n \subket{n+1,n}$, is comprised by states with one extra spin-up sublattice-magnon. (c) Squared amplitudes corresponding to the sublattice-magnon states constituting the uniform squeezed vacuum and spin-up eigenmodes for squeeze parameters of 3 (main) and 1 (inset).}
		\label{fig:main}
	\end{center}
\end{figure}

We here develop the squeezing picture for the ground state and excitations of a simple, two-sublattice AFM. It continuously connects and allows a unified understanding of classical and quantum as well as ordered and disordered antiferromagnetic states. We show that the AFM eigenmodes are obtained by pairwise, two-mode squeezing of sublattice-magnons, the spin-1 excitations delocalized over one of the two sublattices. Focusing on spatially uniform modes, the antiferromagnetic ground state is a superposition of states with equal number of spin-up and -down sublattice-magnons [Fig.~\ref{fig:main}(a) and (c)]. The result is a diminished net spin on each sublattice by an amount dictated by the degree of squeezing, parametrized by the non-negative squeeze parameter $r$. Similarly, a spin-up AFM (squeezed-)magnon is comprised by a superposition of states with $n+1$ spin-up and $n$ spin-down sublattice-magnons [Fig.~\ref{fig:main}(b) and (c)]. Thus, despite its unit net spin, it carries enormous spins on each sublattice which allows it to couple strongly with other excitations via a sublattice-spin mediated interaction (Fig.~\ref{fig:interaction}). Owing to a perfect correlation between the two sublattice-magnon numbers, AFM squeezed-magnons are shown to embody entanglement quantified by von Neumann entropy~\cite{Gerry2004,Nishioka2018} increasing monotonically with $r$ (Fig.~\ref{fig:vNentropy}). The degree of squeezing and entanglement embodied by these eigenmodes is significantly larger than that in hitherto achieved non-equilibrium states. We also comment on existing experiments~\cite{Rodrigue1960,Liensberger2019}, where this squeezing-mediated coupling enhancement (Fig.~\ref{fig:interaction}) has been observed, and strategies for exploiting the entanglement contained in antiferromagnetic magnons. While the squeezed states of light are generated via external drives and are nonequilibrium states~\cite{Gerry2004}, the antiferromagnetic squeezed-magnons are eigenmodes of the system with their squeezing being equilibrium in nature and resulting from energy minimization.

\section{AFM eigenmodes as squeezed Fock states}
We consider a N{\'e}el ordered ansatz with sublattice A and B spins pointing along $\hat{\pmb{z}}$ and $-\hat{\pmb{z}}$, respectively. The antiferromagnetic Hamiltonian may then be expressed in terms of the corresponding sublattice-magnon ladder operators $\tilde{a}_{\pmb{k}}, \tilde{b}_{\pmb{k}}$ as~\cite{Kittel1963,Kamra2017A}:
\begin{align}\label{eq:hamgen}
\tilde{H} = & \sum_{\pmb{k}} A_{\pmb{k}} \left( \tilde{a}^\dagger_{\pmb{k}} \tilde{a}_{\pmb{k}} + \tilde{b}^\dagger_{\pmb{k}} \tilde{b}_{\pmb{k}} \right) + C_{\pmb{k}} \left( \tilde{a}_{\pmb{k}} \tilde{b}_{-\pmb{k}} + \tilde{a}_{\pmb{k}}^\dagger \tilde{b}_{-\pmb{k}}^\dagger \right),
\end{align} 
where we assume inversion symmetry and disregard applied magnetic fields, for simplicity. Consistent with the assumed N{\'e}el order, sublattice B (A) magnons represented by $\tilde{b}_{\pmb{k}}~(\tilde{a}_{\pmb{k}})$ are spin-up (-down). In addition to the general considerations captured by Eq.~(\ref{eq:hamgen}), we will obtain specific results for a uniaxial, easy-axis AFM described by:
\begin{align}\label{eq:hamuni}
\tilde{H}_{\mathrm{uni}} = & ~\frac{J}{\hbar^2} \sum_{i,\pmb{\delta}} \tilde{\pmb{S}}_{\mathrm{A}}(\pmb{r}_i) \cdot \tilde{\pmb{S}}_{\mathrm{B}}(\pmb{r}_i + \pmb{\delta}) \nonumber \\
  & - \frac{K}{\hbar^2} \sum_i \left[ \tilde{S}_{\mathrm{A}z}(\pmb{r}_i) \right]^2 - \frac{K}{\hbar^2} \sum_j \left[ \tilde{S}_{\mathrm{B}z}(\pmb{r}_j) \right]^2.
\end{align}
Here, the positive parameters $J$ and $K$ account for intersublattice antiferromagnetic exchange and easy-axis anisotropy, respectively. $\tilde{\pmb{S}}_{\mathrm{A,B}}$ represent the respective spin operators, $\pmb{r}_{i}~(\pmb{r}_{j})$ runs over the sublattice A (B), and $\pmb{\delta}$ are vectors to the nearest neighbors. Executing Holstein-Primakoff transformations~\cite{Holstein1940} and switching to Fourier space, Eq.~(\ref{eq:hamuni}) reduces to Eq.~(\ref{eq:hamgen}) apart from a constant energy offset~\cite{Akhiezer1968,Kamra2017A}, with $A_{\pmb{k}} = J S z + 2 K S$ and $C_{\pmb{k}} = J S z \gamma_{\pmb{k}}$. Here, $S$ is the spin on each site, $z$ is the coordination number, and $\gamma_{\pmb{k}} \equiv (1/z) \sum_{\pmb{\delta}} \exp \left(i \pmb{k}\cdot \pmb{\delta} \right)$.

The Hamiltonian [Eq.~(\ref{eq:hamgen})] is diagonalized to $\tilde{H} = \sum_{\pmb{k}} \epsilon_{\pmb{k}} \left( \tilde{\alpha}^\dagger_{\pmb{k}} \tilde{\alpha}_{\pmb{k}} + \tilde{\beta}^\dagger_{\pmb{k}} \tilde{\beta}_{\pmb{k}} \right) $ via a Bogoliubov transformation~\cite{Holstein1940} described by~\footnote{We assume $C_{\pmb{k}}$ to be positive.}:
\begin{align}
\tilde{\alpha}_{\pmb{k}} = & ~u_{\pmb{k}} \tilde{a}_{\pmb{k}} + v_{\pmb{k}} \tilde{b}_{-\pmb{k}}^\dagger, \quad \tilde{\beta}_{\pmb{k}} = u_{\pmb{k}} \tilde{b}_{\pmb{k}} + v_{\pmb{k}} \tilde{a}_{-\pmb{k}}^\dagger, \label{eq:magop} \\
u_{\pmb{k}} = & ~\sqrt{\frac{A_{\pmb{k}} + \epsilon_{\pmb{k}}}{2 \epsilon_{\pmb{k}}}},  \ \qquad v_{\pmb{k}} = ~\sqrt{\frac{A_{\pmb{k}} - \epsilon_{\pmb{k}}}{2 \epsilon_{\pmb{k}}}}, \label{eq:uv}
\end{align} 
where $\epsilon_{\pmb{k}} = \sqrt{A_{\pmb{k}}^2 - C_{\pmb{k}}^2}$. $\tilde{\alpha}_{\pmb{k}}$ and $\tilde{\beta}_{\pmb{k}}$ represent the spin-down and -up eigenmodes of the AFM, which are subsequently called squeezed-magnons. Denoting the resulting antiferromagnetic vacuum or ground state wavefunction by $\sqket{\mathrm{G}}$, we have $\tilde{\alpha}_{\pmb{k}} \sqket{\mathrm{G}} = \tilde{\beta}_{\pmb{k}} \sqket{\mathrm{G}} = 0$ for all $\pmb{k}$. 

Let us first consider the spatially uniform modes, i.e. $\pmb{k} = \pmb{0}$. We denote states in the corresponding reduced subspaces via $\subket{N_{b_{\pmb{0}}},N_{a_{\pmb{0}}}}$ and $\sqket{N_{\beta_{\pmb{0}}},N_{\alpha_{\pmb{0}}}}$, where $N_{b_{\pmb{0}}}$ denotes the number of spin-up sublattice-magnons and so on. Within the reduced subspaces, the N{\'e}el ordered state is thus denoted by $\subket{0,0}$, while the antiferromagnetic ground state obtained above is represented by $\sqket{0,0}$. We define the relevant two-mode squeeze operator~\cite{Gerry2004}: $\tilde{S}_{2}(r_{\pmb{0}}) \equiv \exp \left( r_{\pmb{0}} \tilde{a}_{\pmb{0}} \tilde{b}_{\pmb{0}} - r_{\pmb{0}} \tilde{a}_{\pmb{0}}^\dagger \tilde{b}_{\pmb{0}}^\dagger \right)$, with the non-negative squeeze parameter $r_{\pmb{0}}$ given via $u_{\pmb{0}} \equiv \cosh r_{\pmb{0}}$ and $v_{\pmb{0}} \equiv \sinh r_{\pmb{0}}$ [Eq.~(\ref{eq:uv})]~\footnote{In defining the squeeze operator, we have implicitly assumed positive $C_{\pmb{k}}$. If $C_{\pmb{k}}$ is negative, we obtain the same non-negative squeeze parameter with a squeezing phase of $\pi$~\cite{Gerry2004}. The phenomena studied herein remain unaffected under such a phase shift.}. Employing the identities~\cite{Gerry2004,Kamra2016A}:
\begin{align} 
\tilde{\alpha}_{\pmb{0}} & = \tilde{S}_{2}(r_{\pmb{0}}) \tilde{a}_{\pmb{0}} \tilde{S}_{2}^{-1}(r_{\pmb{0}}), \quad \tilde{\beta}_{\pmb{0}} = \tilde{S}_{2}(r_{\pmb{0}}) \tilde{b}_{\pmb{0}} \tilde{S}_{2}^{-1}(r_{\pmb{0}}),
\end{align}
where $\tilde{\alpha}_{\pmb{0}}$ and $\tilde{\beta}_{\pmb{0}}$ are given by Eq.~(\ref{eq:magop}), into the condition $\tilde{\alpha}_{\pmb{0}} \sqket{0,0} = \tilde{\beta}_{\pmb{0}} \sqket{0,0} = 0$, we obtain:
\begin{align}
\sqket{0,0} = \tilde{S}_{2}(r_{\pmb{0}}) \subket{0,0}.
\end{align} 
Thus, the uniform modes antiferromagnetic ground state is a two-mode squeezed vacuum of sublattice-magnons. The complementary demonstration of quadrature squeezing has been detailed in Appendix \ref{AppA}. Working along the same lines as above, it is straightforward to show that $\sqket{m,n} = \tilde{S}_{2}(r_{\pmb{0}}) \subket{m,n}$, thereby demonstrating the antiferromagnetic eigenmodes to be two-mode squeezed sublattice-magnon Fock states. Therefore, the eigenmodes are henceforth called ``squeezed-magnons''. 

Based on the analysis above, it becomes evident that the antiferromagnetic ground state is obtained by pairwise, two-mode squeezing of the N{\'e}el ordered state:
\begin{align}
\sqket{\mathrm{G}} = & \left[ \prod_{\pmb{k}} \tilde{S}_{2} \left( r_{\pmb{k}}\right)  \right] \subket{\mathrm{N\acute{e}el}},
\end{align}
where $\tilde{S}_{2} \left( r_{\pmb{k}} \right) \equiv \exp \left( r_{\pmb{k}} \tilde{a}_{\pmb{k}} \tilde{b}_{-\pmb{k}} - r_{\pmb{k}} \tilde{a}_{\pmb{k}}^\dagger \tilde{b}_{-\pmb{k}}^\dagger \right)$, with the squeeze parameters $r_{\pmb{k}}$ given via $u_{\pmb{k}} = u_{-\pmb{k}} \equiv \cosh r_{\pmb{k}} $. The $\tilde{\alpha}_{\pmb{k}}$ eigenmode is thus a two-mode ($\tilde{a}_{\pmb{k}}$ and $\tilde{b}_{-\pmb{k}}$) squeezed-magnon [Eq. (\ref{eq:magop})]. Similarly, the $\tilde{\beta}_{\pmb{k}}$ eigenmode is also a two-mode squeezed magnon formed by $\tilde{b}_{\pmb{k}}$ and $\tilde{a}_{-\pmb{k}}$ modes [Eq. (\ref{eq:magop})]. Due to this mathematical equivalence, it suffices to analyze the spatially uniform eigenmodes, which is what we focus on in the following.

\section{Spatially uniform eigenmodes}

For ease of notation, we denote the wavefunctions for spatially uniform squeezed vacuum by $\sqket{0}$ and spin-up squeezed-magnon by $\sqket{\uparrow}$, while the corresponding squeeze parameter is denoted by $r$. Considering a uniaxial AFM [Eq.~(\ref{eq:hamuni})], we obtain $\cosh r \approx (1/2) (J z/K)^{1/4}$ [Eq.~(\ref{eq:uv})], which translates to $r \approx 3$ for a typical ratio of $J/K \sim 10^4$. To get a feel for numbers, the most squeezed vacuum state of light generated so far corresponds to a squeeze parameter of about 1.7~\cite{Vahlbruch2016,Schnabel2017}. Furthermore, in the limit $K \to 0$, the squeeze parameter is found to diverge. This feature is general and a direct consequence [Eq.~(\ref{eq:uv})] of the Goldstone theorem, according to which $\epsilon_{\pmb{0}} \to 0$ in the limit of isotropy. 

Employing the relation $\tilde{\alpha}_{\pmb{0}} \sqket{0} = (\cosh r ~\tilde{a}_{\pmb{0}} + \sinh r~ \tilde{b}_{\pmb{0}}^\dagger) \sqket{0} = 0$, the squeezed vacuum is obtained in terms of the uniform sublattice-magnons subspace~\cite{Gerry2004}:
\begin{align}\label{eq:pn}
\sqket{0} = & \sum_{n=0}^{\infty} \frac{\left(- \tanh r \right)^n}{\cosh r} \subket{n,n} \equiv \sum_{n} P_n \subket{n,n}. 
\end{align} 
The ensuing wavefunction is schematically depicted in Fig.~\ref{fig:main}(a) and the distribution over constituent states is plotted in Fig.~\ref{fig:main}(c). With an increasing $r$, the number of states that contribute substantially to the superposition increases monotonically. This presence of sublattice-magnons in the ground state constitutes quantum fluctuations. 

\begin{figure}[tb]
	\begin{center}
		\subfloat[]{\includegraphics[height=30mm]{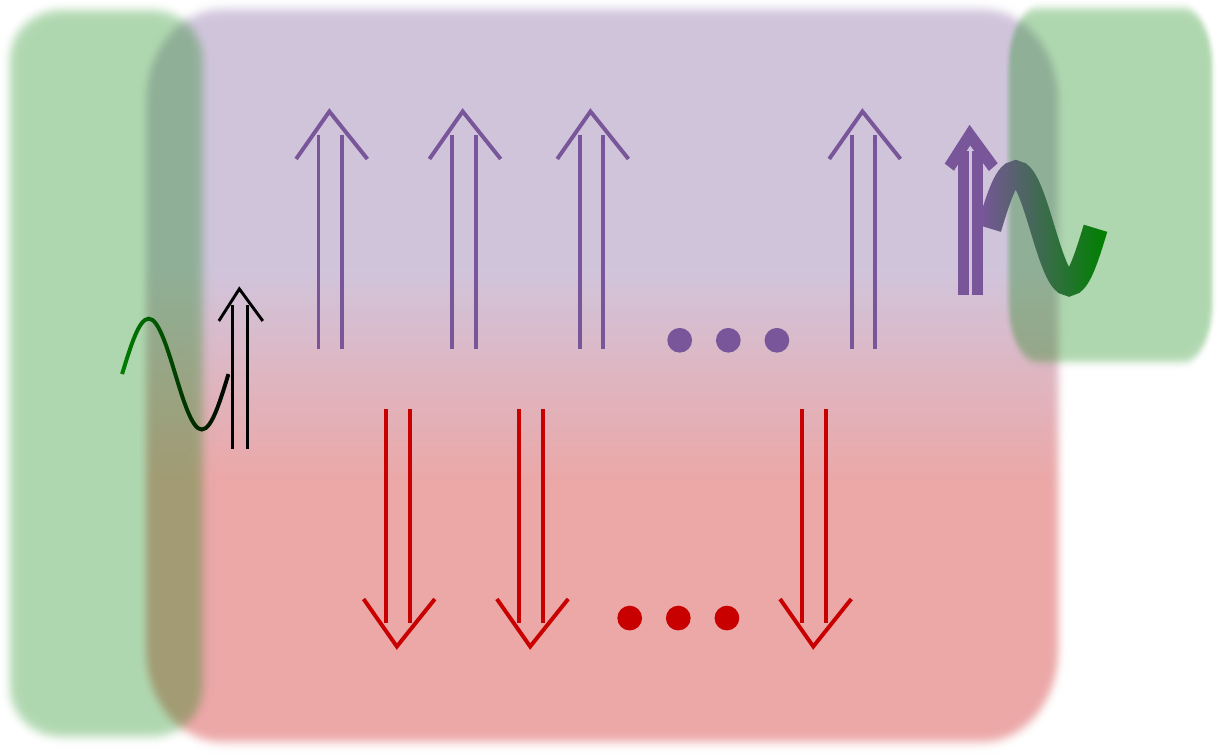}} \qquad 
		\subfloat[]{\includegraphics[height=30mm]{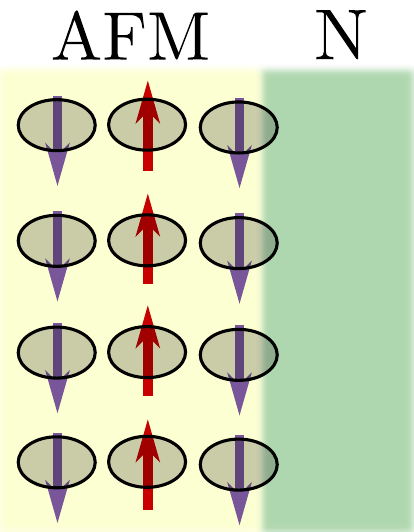}}
		\caption{(a) An external excitation bath (shaded green) interacts weakly with the AFM squeezed-magnon if coupled via its unit net spin (left), but strongly if exposed to only one of the sublattices (right). (b) Schematic depiction of a metal (N) coupled to an AFM via a fully uncompensated interface.}
		\label{fig:interaction}
	\end{center}
\end{figure}

A similar representation for the spin-up squeezed-magnon is obtained via $\sqket{\uparrow} = \tilde{\beta}_{\pmb{0}}^\dagger \sqket{0} = (\cosh r ~\tilde{b}_{\pmb{0}}^\dagger + \sinh r~ \tilde{a}_{\pmb{0}}) \sqket{0}$ and Eq.~(\ref{eq:pn}):
\begin{align}\label{eq:qn}
\sqket{\uparrow} = &  \sum_{n=0}^{\infty} \frac{ \sqrt{n+1} \left(- \tanh r \right)^n}{\cosh^2 r} \subket{n+1,n}, \nonumber \\
  & \equiv \sum_{n} Q_n \subket{n+1,n}.
\end{align} 
A schematic depiction and the distribution over constituent states are shown in Fig.~\ref{fig:main}(b) and (c). In stark contrast with the squeezed vacuum, where the contribution from states decreases monotonically with $n$, the highest contribution to the superposition here comes from $n \approx \sinh^2 r$. No such peak exists for weak squeezing when $\sinh r < 1$. The average number of spin-up magnons comprising a squeezed-magnon is evaluated as $\cosh^2 r + \sinh^2 r$. Thus, a typical AFM squeezed-magnon, corresponding to $r \approx 3$ estimated above, is comprised by around 200 spin-up magnons on one sublattice and nearly the same number of spin-down magnons on the other. It is thus an enormous excitation, despite its unit net spin. 


\section{Enhanced interaction}

This enormous nature of the AFM squeezed-magnon reveals an approach to exploit it. When it couples to excitations, such as itinerant electrons or phonons, via its net spin, the interaction strength is proportional to the relatively small unit spin. On the other hand, if an interaction is mediated via the sublattice-spin, it will be greatly enhanced (by a factor $\sim  \cosh^2 r \approx 100$ for $r \approx 3$) on account of its large sublattice spin content [Fig.~\ref{fig:interaction}(a)]. Such a situation arises, for example, when an AFM is exposed to a metal via an uncompensated interface [Fig.~\ref{fig:interaction}(b)]~\cite{Manna2014,Zhang2016,Kamra2017B,Kamra2018A}. This effect provides a physical picture for the theoretically encountered enhancement in spin pumping current from AFM into an adjacent conductor coupled asymmetrically to the two sublattices~\cite{Kamra2017B}. The same mechanism has also been exploited in predicting an enhanced magnon-mediated superconductivity in a conductor bearing an uncompensated interface with an AFM~\cite{Erlandsen2019}. Rigorous derivations of electron-magnon and magnon-magnon couplings presented respectively in Appendices \ref{AppB} and \ref{AppC} demonstrate an enhancement in the interactions consistent with the intuition above reinforcing the generality of this phenomenon.


\begin{figure}[tb]
	\begin{center}
		\includegraphics[width=70mm]{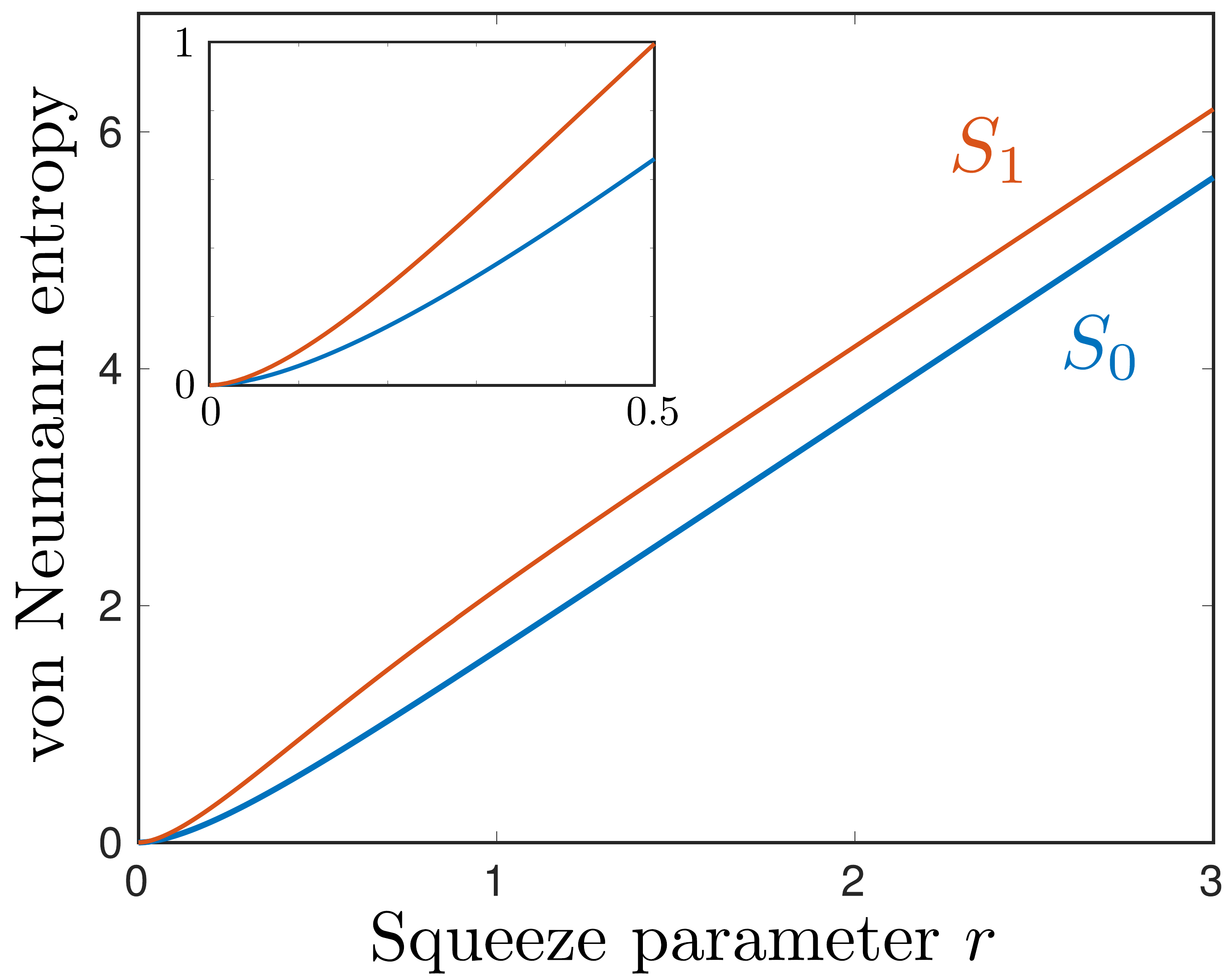}
		\caption{Entanglement between the two constituent sublattice-magnons quantified via von Neumann entropy for the squeezed vacuum ($S_0$) and magnon ($S_1$) eigenmodes. The inset shows a zoom-in of the small $r$ range.} 
		\label{fig:vNentropy}
	\end{center}
\end{figure}

\section{Entanglement}

In a two-mode squeezed vacuum, the participating modes are entangled with the degree of entanglement quantified by the von Neumann entropy~\cite{Nishioka2018,Gerry2004} $S_{0}$:
\begin{align}
S_{0} & = ~ - \sum_{n} |P_n|^2 \ln \left( |P_n|^2 \right), \nonumber \\
  & = ~ 2 \ln (\cosh r ) - 2 (\sinh^2 r) \ln (\tanh r). 
\end{align}
Such two-mode squeezed vacuum states of light have been exploited for obtaining useful entanglement~\cite{Ou1992}. This high von Neumann entropy content of our squeezed-magnon vacuum can be exploited, for example, in entangling two qubits~\cite{Zou2019} coupled respectively to sublattices A and B. Furthermore, the squeezed-magnons themselves embody strong entanglement, quantified by an even larger von Neumann entropy $S_{1} = - \sum_{n} |Q_n|^2 \ln ( |Q_n|^2 )$ (Fig.~\ref{fig:vNentropy}), which may be transfered to external excitations. This can be achieved by coupling the systems to be entangled with the opposite sublattices~\cite{Cornelissen2015,Goennenwein2015,Bender2019,Johansen2019,MacNeill2019}, via uncompensated interfaces [Fig.~\ref{fig:interaction}(b)], for example, as has been detailed further in Appendix \ref{AppD}. In comparison, von Neumann entropy~\footnote{Strictly speaking, second-order R{\'e}nyi entropy, which provides a lower bound on von Neumann entropy, was measured.} of about 1 has been measured in cold atom systems~\cite{Islam2015}. This high von Neumann entropy content and the large number of entangled spins ($\sim \cosh^2 r $) that comprise the AFM squeezed-magnon make it an entangled excitation complementary to the ``massively entangled'' excitations hosted by some quantum spin liquids~\cite{Castelnovo2008,Balents2010,Savary2016}.


\section{Quantum fluctuations in ``classical'' experiments}
	
The interaction enhancement effect [Fig.~\ref{fig:interaction}(a)] is rooted in high magnon-squeezing and the underlying quantum superposition of a large number of states [Eq.~(\ref{eq:qn})]. It is a direct consequence of the strong quantum fluctuations in the antiferromagnetic ground state, that hosts this excitation, and is thus a quantum fluctuation effect itself. Nevertheless, this coupling enhancement is observed as an increased magnetic damping around compensation temperature in a compensated ferrimagnet~\cite{Rodrigue1960}, which mimics an AFM~\cite{Kamra2017A,Kamra2018B}. Recently, this enhancement has been observed and exploited in a compensated ferrimagnet for an ultrastrong magnon-magnon coupling resulting in hybridization between the two enormous spin-up and -down squeezed-magnons~\cite{Liensberger2019}. These ``classical'' experiments at high temperatures may thus be considered observation of the antiferromagnetic quantum fluctuations. As detailed in Appendix \ref{AppC}, this high squeezing-mediated enhancement ($\sim \sqrt{J/K}$ for our uniaxial AFM), suggested recently in the context of light-matter interaction~\cite{Leroux2018,Qin2018}, is reproduced by the classical theory of spin dynamics~\cite{Kamra2018B,Liensberger2019}, where it is termed ``exchange-enhancement''. This is understandable since the classical dynamics is captured by the quantum system being in a coherent state~\cite{Glauber1963,Sudarshan1963,Kamra2017B}, which fully accounts for the average effect of these quantum fluctuations.

\section{Generalizations}

The description in terms of squeezed Fock states developed herein is a mathematical consequence of the Bogoliubov transformation and goes beyond AFMs. It should allow a similar physical picture, and subsequent exploitation of quantum effects, in other systems such as cold atoms~\cite{Bloch2008,Galitski2013,Galitski2019}. Here, we have disregarded the relatively weak spin-nonconserving interactions. Inclusion of those necessitates a 4-dimensional Bogoliubov transform~\cite{Kamra2017A} thereby precluding the simple two-mode squeezed Fock states description employed here. Similar complications also arise when considering AFMs lacking inversion symmetry. Nevertheless, an analogous general picture can be developed.

\section{Conclusion}

We have developed a description and physical picture of antiferromagnetic ground state and excitations based on the concept of two-mode squeezed Fock states. Capitalizing on the tremendous progress in quantum optics, these fresh insights pave the way for exploiting the quantum properties of antiferromagnetic squeezed-magnons towards, potentially room temperature, quantum devices.   

\section*{Acknowledgments}
A.K.~thanks So Takei, Lukas Liensberger, Mathias Weiler, and Hans Huebl for valuable discussions. We acknowledge financial support from the Research Council of Norway through its Centers of Excellence funding scheme, project 262633, ``QuSpin'', and the DFG through SFB 767. A.S.~also acknowledges support from the Research Council of Norway, grant No. 250985, ``Fundamentals of Low-dissipative Topological Matter''.

\appendix


\section{Demonstration of Quadrature Squeezing}\label{AppA}
In this section, we clarify the squeezed nature of the antiferromagnetic ground state by evaluating the quantum fluctuations in the appropriate quadratures. This approach is complementary to the more general discussion in terms of the two-mode squeeze operator~\cite{Gerry2004} presented in the main text. Once again, we focus on the uniform modes, i.e. $\pmb{k} = \pmb{0}$, recognizing that the corresponding results for $\pmb{k} \neq \pmb{0}$ follow in a similar fashion. We first demonstrate the quadrature squeezing following the standard approach within quantum optics~\cite{Gerry2004} and physically interpret the quadratures later.

For the two-mode squeezing of $\tilde{a}_{\pmb{0}}$ and $\tilde{b}_{\pmb{0}}$ operational here, the relevant quadratures are formed via a combination of both modes' ladder operators~\cite{Gerry2004}:
\begin{align}
\tilde{X}_1 & \equiv  \frac{1}{\sqrt{8}} \left( \tilde{a}_{\pmb{0}} + \tilde{a}_{\pmb{0}}^\dagger + \tilde{b}_{\pmb{0}} + \tilde{b}_{\pmb{0}}^\dagger \right), \label{eq:x1} \\
\tilde{X}_2 & \equiv  \frac{1}{i \sqrt{8}} \left( \tilde{a}_{\pmb{0}} - \tilde{a}_{\pmb{0}}^\dagger + \tilde{b}_{\pmb{0}} - \tilde{b}_{\pmb{0}}^\dagger \right). \label{eq:x2}
\end{align}
Employing the bosonic commutation relations of the ladder operators, we obtain $[ \tilde{X}_1 , \tilde{X}_2 ] = i/2$, demonstrating that the chosen quadratures of Eqs. (\ref{eq:x1}) and (\ref{eq:x2}) represent two noncommuting observables. Denoting the reduced subspace of the uniform modes within the N{\'e}el ordered state by $\subket{0}$, the quantum fluctuations in the two quadratures are evaluated as:
\begin{align}
\subbra{0} (\delta \tilde{X}_1)^2 \subket{0} & \equiv \subbra{0} (\tilde{X}_1 - \langle \tilde{X}_1 \rangle )^2 \subket{0} = \frac{1}{4} , \\
\subbra{0} (\delta \tilde{X}_2 )^2 \subket{0} & =  \frac{1}{4}.
\end{align}
Therefore the two quadratures host equal quantum noise in the N{\'e}el ordered state, that is $\subbra{0} (\delta \tilde{X}_1)^2 \subket{0} = \subbra{0} (\delta \tilde{X}_2)^2 \subket{0}$.

We now consider fluctuations in the antiferromagnetic ground state with the uniform modes reduced subspace denoted by $\sqket{0}$, as in the main text. Employing the Bogoliubov transformation relations $\tilde{a}_{\pmb{0}} = \cosh r ~ \tilde{\alpha}_{\pmb{0}} - \sinh r ~ \tilde{\beta}_{\pmb{0}}^\dagger $ and $\tilde{b}_{\pmb{0}} = \cosh r ~ \tilde{\beta}_{\pmb{0}} - \sinh r ~ \tilde{\alpha}_{\pmb{0}}^\dagger $, the two quadratures can be expressed as:
\begin{align}
\tilde{X}_1 = & \frac{\cosh r - \sinh r}{\sqrt{8}} \left( \tilde{\alpha}_{\pmb{0}} + \tilde{\alpha}_{\pmb{0}}^\dagger + \tilde{\beta}_{\pmb{0}} + \tilde{\beta}_{\pmb{0}}^\dagger \right), \\
\tilde{X}_2 = & \frac{\cosh r + \sinh r}{i \sqrt{8}} \left( \tilde{\alpha}_{\pmb{0}} - \tilde{\alpha}_{\pmb{0}}^\dagger + \tilde{\beta}_{\pmb{0}} - \tilde{\beta}_{\pmb{0}}^\dagger \right).
\end{align}
Employing the quadrature expressions thus obtained, quantum fluctuations in the antiferromagnetic ground state are conveniently evaluated as:
\begin{align}
\sqbra{0} (\delta \tilde{X}_1)^2 \sqket{0} = &  \frac{\left( \cosh r - \sinh r \right)^2}{4} = \frac{e^{-2r}}{4} , \\
\sqbra{0} (\delta \tilde{X}_2 )^2 \sqket{0} = & \frac{\left( \cosh r + \sinh r \right)^2}{4} = \frac{e^{2r}}{4},
\end{align}
thereby demonstrating the quadrature squeezing~\cite{Gerry2004} of the antiferromagnetic ground state, that is $\sqbra{0} (\delta \tilde{X}_1 )^2 \sqket{0} < \sqbra{0} (\delta \tilde{X}_2 )^2 \sqket{0}$. 

We now relate the two quadratures [Eqs. (\ref{eq:x1}) and (\ref{eq:x2})] with physical observables of the antiferromagnet (AFM). Employing Fourier relations of the kind
\begin{align}
\tilde{a}_{\pmb{k}} = & \frac{1}{\sqrt{N}} \sum_i \tilde{a}_i ~ e^{i \pmb{k}\cdot \pmb{r}_i} ,
\end{align}
in conjunction with the linearized Holstein-Primakoff transformations for the AFM~\cite{Akhiezer1968,Kittel1963}:
\begin{align}
\tilde{S}_{A+}(\pmb{r}_i) = \tilde{S}_{Ax}(\pmb{r}_i) + i \tilde{S}_{Ay}(\pmb{r}_i) &=   \hbar \sqrt{2 S} ~ \tilde{a}_i, \\
\tilde{S}_{B+}(\pmb{r}_j) = \tilde{S}_{Bx}(\pmb{r}_j) + i \tilde{S}_{By}(\pmb{r}_j) & =   \hbar \sqrt{2 S} ~ \tilde{b}_j^\dagger,
\end{align}
we obtain
\begin{align}
\tilde{X}_1 & =  \frac{1}{2 \hbar \sqrt{NS}} \left(\tilde{S}_{Ax} + \tilde{S}_{Bx}\right), \label{eq:x1phys}\\
\tilde{X}_2 & =  \frac{1}{2 \hbar \sqrt{NS}} \left(\tilde{S}_{Ay} - \tilde{S}_{By}\right). \label{eq:x2phys}
\end{align}
Here, $N$ is the total number of sites on each sublattice, $S$ is the spin at each site as defined in the main text, and $\tilde{S}_{Ax} \equiv \sum_i \tilde{S}_{Ax}(\pmb{r}_i)$ is the x component of the total spin on sublattice A, and so on. Thus, the two quadratures are related to the x and y components of the total spin and the N{\'e}el order, respectively. 

In the qualitatively distinct case of single-mode squeezing manifested by the uniform mode in an anisotropic ferromagnet~\cite{Kamra2016A}, the two quadratures are simply the x and y components of the total spin providing a geometrical ``ellipticity'' interpretation to the squeezing effect~\footnote{Quadrature squeezing however comments on the ellipticity in the quantum fluctuations and not the expectation values of the spins in the coherent state, as is the case with the ellipticity in classical spin wave picture. The two kinds of ellipticities, although interrelated in equilibrium for the case under discussion, do not need to be identical in general.}. In contrast, the situation is less intuitive for the case of two-mode squeezing as the ellipticity of quantum fluctuations exists in a more abstract space. In the present case, this space is defined by the transverse orthogonal components of the total spin and the N{\'e}el order associated with the AFM [Eqs. (\ref{eq:x1phys}) and (\ref{eq:x2phys})].


\section{Electron-magnon coupling}\label{AppB}

Heterostructures in which a magnetic insulator layer is interfaced with another material hosting conduction electrons have emerged as basic building blocks in a wide range of spintronic concepts and devices. The interfacial exchange-mediated coupling between the magnons in the former and the electrons in the latter have enabled magnon-based information processing schemes, magnon-mediated condensation phenomena and so on. Thus, an ability to engineer and amplify the electron-magnon coupling is expected to have a strong and broad impact. In this section, we discuss the electron-magnon coupling in an AFM/normal metal (N) bilayer with the goal of highlighting this tunability and amplification of electron-magnon coupling by exploiting the squeezing effect, as discussed in the main text. A thorough analysis of this system along with spin transport effects has been provided elsewhere~\cite{Kamra2017B}. We here focus on highlighting the amplification effect for an uncompensated AFM with respect to other related systems, providing mathematical expressions complementary to the intuitive physical picture discussed in the main text.

The AFM and N layers are assumed to interact via interfacial exchange resulting in the following contribution to the Hamiltonian~\cite{Kamra2017B} within a continuum model:
\begin{align}\label{eq:hint1}
\tilde{H}_{\mathrm{int}} & = - \frac{1}{\hbar^2} \int_{\mathcal{A}} d^2 \rho \sum_{\mathrm{G} = \mathrm{A},\mathrm{B}} \mathcal{J}_{\mathrm{iG}} ~ \tilde{S}_{\mathrm{G}} (\pmb{\rho}) \cdot \tilde{S}_{\mathrm{N}}(\pmb{\rho}),
\end{align} 
where $\mathcal{A}$ is the interfacial area, $\pmb{\rho}$ is the two-dimensional position vector in the interfacial plane, $\tilde{S}_{\mathrm{N}}$ is the conduction electrons spin density operator in N, $\tilde{S}_{\mathrm{G}}$ is the spin density operator in the magnet for sublattice G, and $\mathcal{J}_{\mathrm{iG}}$ parametrizes the exchange interaction between the two spin densities allowing it to be sublattice asymmetric. In terms of the ladder operators for the conduction electrons and magnons, the Hamiltonian above takes the form:
\begin{align}\label{eq:hint2}
\tilde{H}_{\mathrm{int}} & = \hbar \sum_{\pmb{q}_1,\pmb{q}_2,\pmb{k}} \tilde{c}_{\pmb{q}_1 +}^\dagger \tilde{c}_{\pmb{q}_2 -}  \left( W^{\mathrm{A}}_{\pmb{q}_1 \pmb{q}_2 \pmb{k}} \tilde{a}_{\pmb{k}}  + W^{\mathrm{B}}_{\pmb{q}_1 \pmb{q}_2 \pmb{k}} \tilde{b}_{\pmb{k}}^\dagger \right)   + \mathrm{h.c.},
\end{align} 
where $\tilde{c}_{\pmb{q} +}$ denotes the annihilation operator for the N conduction electron with wavevector $\pmb{q}$ and spin $+ \hbar/2$ along the z-direction and so on, $\tilde{a}_{\pmb{k}}$ and $\tilde{b}_{\pmb{k}}$ are the annihilation operators for the sublattice-magnons as discussed in the main text, $ W^{\mathrm{A}}_{\pmb{q}_1 \pmb{q}_2 \pmb{k}}$ is the appropriate amplitude given by the overlap integral between the participating excitation wavefunctions~\cite{Kamra2017B}. With the aim of focusing on the key ingredient in enhancing the coupling, we henceforth consider the relevant and simplified part of the Hamiltonian [enclosed by brackets in Eq. (\ref{eq:hint2})] describing electron-magnon coupling:
\begin{align}\label{eq:p}
\tilde{P} & =  W^{\mathrm{A}} \tilde{a}_{\pmb{0}}  + W^{\mathrm{B}} \tilde{b}_{\pmb{0}}^\dagger ,
\end{align}
where we have again specialized the expression to uniform ($\pmb{k} = \pmb{0}$) modes for simplicity, $W^{\mathrm{A,B}} \propto \mathcal{J}_{\mathrm{iA,iB}}$ capture the sublattice-asymmetry in the interfacial coupling.  

For comparison, we first consider the case of a single-sublattice isotropic ferromagnet~\cite{Kamra2016A} for which the interaction is described simply by $\tilde{P} = W \tilde{a}_{\pmb{0}}$, with $\tilde{a}_{\pmb{0}}$ representing the normal magnon mode. The transition rate $\Gamma$ for the electron-magnon scattering process is thus simply determined by $W$, i.e. $\Gamma \propto |W|^2$. For the case of AFMs, in contrast, Eq. (\ref{eq:p}) becomes
\begin{align}
\tilde{P} = & \left( \cosh r ~W^{\mathrm{A}} - \sinh r ~W^{\mathrm{B}} \right) \tilde{\alpha}_{\pmb{0}} + \nonumber \\
 & \left( \cosh r ~W^{\mathrm{B}} - \sinh r~ W^{\mathrm{A}} \right) \tilde{\beta}_{\pmb{0}}^\dagger,
\end{align}
in terms of the normal magnon modes. Now considering $W^{\mathrm{A}} = W^{\mathrm{B}} \equiv W$ for a compensated interface, in which the two sublattices couple equally to the N electrons, we obtain:
\begin{align}
\tilde{P} & =   W \left( \cosh r  - \sinh r  \right) \tilde{\alpha}_{\pmb{0}} + W \left( \cosh r  - \sinh r \right)  \tilde{\beta}_{\pmb{0}}^\dagger,
\end{align}
whence we see that the transition rate is reduced: $\Gamma \propto  \left( \cosh r  - \sinh r  \right)^2 |W|^2 \approx |W|^2 / (4 \cosh^2 r) $, accounting for the large squeezing such that $\cosh r \gg 1$. The electron-magnon coupling for this case is thus suppressed as compared to that for ferromagnetic magnons considered above. Arriving at the crux of this section, as discussed in the main text, when the coupling is mediated by the sublattice-spin of the magnon via an uncompensated interface ($W^\mathrm{A} = W$, $W^\mathrm{B} = 0$), we obtain
\begin{align}
\tilde{P} & =  W  \cosh r ~ \tilde{\alpha}_{\pmb{0}} - W  \sinh r ~ \tilde{\beta}_{\pmb{0}}^\dagger.
\end{align}
The transition rates for the electron-magnon scattering processes are thus given by $\Gamma \propto \cosh^2 r |W|^2$ for $\tilde{\alpha}_{\pmb{0}}$ mode and  $\Gamma \propto \sinh^2 r |W|^2 \approx \cosh^2 r |W|^2$ for the  $\tilde{\beta}_{\pmb{0}}$ mode. Thus, we find a squeezing-mediated enhancement in the electron-magnon coupling for the case of sublattice spin-mediated interaction. Furthermore, this is consistent with the simple picture discussed in the main text and the interaction enhancement factor is related to the sublattice-spin associated with a single eigenexcitation - antiferromagnetic squeezed-magnon.


\section{Magnon-Magnon coupling}\label{AppC}

In this section, we investigate coupling between the two opposite-spin antiferromagnetic eigenmodes caused by a spin-nonconserving interaction~\cite{Kamra2017A}. In particular, we demonstrate that a sublattice spin-mediated magnon-magnon coupling is amplified via the squeezing effect in consistence with the general picture discussed in the main text. This also provides a derivation, within the quantum picture, for the recently observed ``exchange-enhanced'' ultrastrong magnon-magnon coupling in a compensated ferrimagnet~\cite{Liensberger2019} without accounting for all the experimental complexities therein.

In the main text, we have only considered interactions that conserve the z-projected spin of the AFM. The diagonalized Hamiltonian therefore assumes the form:
\begin{align}\label{eq:hmain}
\tilde{H} & = \sum_{\pmb{k}} \epsilon_{\pmb{k}} \left( \tilde{\alpha}_{\pmb{k}}^\dagger \tilde{\alpha}_{\pmb{k}} + \tilde{\beta}_{\pmb{k}}^\dagger \tilde{\beta}_{\pmb{k}}  \right), 
\end{align}
with the two opposite-spin squeezed-magnons as degenerate excitations of the system, in the absence of an applied field. However, breaking the spin conservation~\footnote{In the following discussion, we are concerned with the z-projected spin without specifying this directional preference explicitly.} in the system allows to couple these opposite-spin excitations resulting in a lifting of degeneracy and the concomitant hybridization~\cite{Kamra2017A}. As discussed in the main text, accounting for such spin-nonconserving terms necessitates a four-dimensional Bogoliubov transform for an exact diagonalization of the Hamiltonian~\cite{Kamra2017A}. Here, we circumvent this mathematical complexity by describing the mode-coupling in a perturbative manner treating Eq. (\ref{eq:hmain}) and squeezed-magnons as our unperturbed Hamiltonian and eigenexcitations, respectively. This allows us to obtain an analytic expression for the coupling rate while appreciating and justifying the typical approximations employed in such descriptions~\cite{Gerry2004}.

For concreteness, we consider the following spin-nonconserving and sublattice spin-mediated contribution to the Hamiltonian that may stem from the magnetocrystalline anisotropy~\cite{Liensberger2019}:
\begin{align}\label{eq:hcoup1}
\tilde{H}_{\mathrm{coup}} & = \frac{K_{a}}{\hbar^2} \sum_{i}  \left( \tilde{S}_{\mathrm{Ax}} (\pmb{r}_i) \right)^2 + \frac{K_{a}}{\hbar^2} \sum_{j} \left( \tilde{S}_{\mathrm{Bx}} (\pmb{r}_j) \right)^2, 
\end{align}
where $K_a$ parametrizes this axial-symmetry-breaking anisotropy, and rest of the notation has already been introduced in the main text. Employing Holstein-Primakoff transformation and switching to Fourier space, the coupling Hamiltonian above is brought to the following form:
\begin{align}\label{eq:hcoup2}
\tilde{H}_{\mathrm{coup}} & = \frac{K_a S}{2} \sum_{\pmb{k}}  \tilde{a}_{\pmb{k}}^\dagger \tilde{a}_{- \pmb{k}}^\dagger + \tilde{b}_{\pmb{k}}^\dagger \tilde{b}_{- \pmb{k}}^\dagger + \tilde{a}_{\pmb{k}} \tilde{a}_{- \pmb{k}} + \tilde{b}_{\pmb{k}} \tilde{b}_{- \pmb{k}} . 
\end{align}
In writing Eq. (\ref{eq:hcoup2}) above, we have neglected terms of the type $\sim \tilde{a}_{\pmb{k}}^\dagger \tilde{a}_{\pmb{k}}$ since they can be absorbed into Eq. (\ref{eq:hmain}) leading to a small renormalization of the unperturbed squeezed-magnon energies. We again focus on the uniform modes ($\pmb{k} = \pmb{0}$) as they are also the ones observed experimentally~\cite{Liensberger2019}:
\begin{align}
\tilde{H}_{\mathrm{coup}} (\pmb{k} = \pmb{0}) & = \frac{K_a S}{2} \left( \tilde{a}_{\pmb{0}}^2 + \tilde{b}_{\pmb{0}}^2 + \mathrm{h.c.} \right).
\end{align}  
Employing the Bogoliubov transformation relations $\tilde{a}_{\pmb{0}} = \cosh r ~ \tilde{\alpha}_{\pmb{0}} - \sinh r ~ \tilde{\beta}_{\pmb{0}}^\dagger $ and $\tilde{b}_{\pmb{0}} = \cosh r ~ \tilde{\beta}_{\pmb{0}} - \sinh r ~ \tilde{\alpha}_{\pmb{0}}^\dagger $, the coupling Hamiltonian may be expressed in terms of the unperturbed eigenexcitations:
\begin{align}
\tilde{H}_{\mathrm{coup}} (\pmb{k} = \pmb{0})  & =  - \cosh r \sinh r ~ 2 K_a S \left( \tilde{\alpha}_{\pmb{0}} \tilde{\beta}_{\pmb{0}}^\dagger + \tilde{\alpha}_{\pmb{0}}^\dagger \tilde{\beta}_{\pmb{0}} \right) \nonumber \\
& \quad ~  + \frac{K_a S \left( \cosh^2 r + \sinh^2 r \right)}{2} \left( \tilde{\alpha}_{\pmb{0}}^2 + \tilde{\beta}_{\pmb{0}}^2 + \mathrm{h.c.} \right),  \\
& \approx - \cosh r \sinh r ~ 2 K_a S \left( \tilde{\alpha}_{\pmb{0}} \tilde{\beta}_{\pmb{0}}^\dagger + \tilde{\alpha}_{\pmb{0}}^\dagger \tilde{\beta}_{\pmb{0}} \right) \label{eq:hcoupfin}.
\end{align}  
In the last simplification above, we have employed the rotating wave approximation~\cite{Gerry2004} and disregarded terms which merely cause rapid oscillations. 

Equation (\ref{eq:hcoupfin}) above constitutes the main result of this section whence the coupling rate can be read off as $\cosh r \sinh r ~ 2 K_a S$. The squeezing-mediated enhancement in coupling of $\cosh r \sinh r \approx \cosh^2 r \sim \sqrt{J/K}$ is evident and consistent with the intuitive picture presented in the main text. In comparison, if we consider a net spin-mediated magnon-magnon coupling via, for example,
\begin{align}
\tilde{H}_{\mathrm{coup}} & = \frac{K_{a}}{\hbar^2} \sum  \left( \tilde{S}_{\mathrm{Ax}} (\pmb{r}_i) + \tilde{S}_{\mathrm{Bx}} (\pmb{r}_j) \right)^2,
\end{align}
an analogous procedure yields a suppressed coupling rate of $K_a S / (4 \cosh^2 r)$, in consistence with the electron-magnon coupling considerations discussed above. 

Thus, these two instances (electron-magnon and magnon-magnon couplings) of detailed calculations reinforce the generality of the intuitive picture discussed in the main text. This also suggests these coupling properties to be intrinsic to the antiferromagnetic squeezed-magnons, and therefore applicable to a yet wider class of phenomena involving antiferromagnets. We further note that the squeezing-mediated coupling enhancement that we describe here is mathematically analogous to similar nonequilibrium enhancements suggested recently in the context of light-matter interaction~\cite{Leroux2018,Qin2018}. Our suggestion for magnets bears advantages such as stronger enhancement, equilibrium nature of the effect, tunability via temperature~\cite{Liensberger2019}, and the recent experimental observation~\cite{Liensberger2019} along with the concomitant proof-of-concept.


\section{Accessing entangled subsystems}\label{AppD}

\begin{figure}[tb]
	\begin{center}
		\includegraphics[width=50mm]{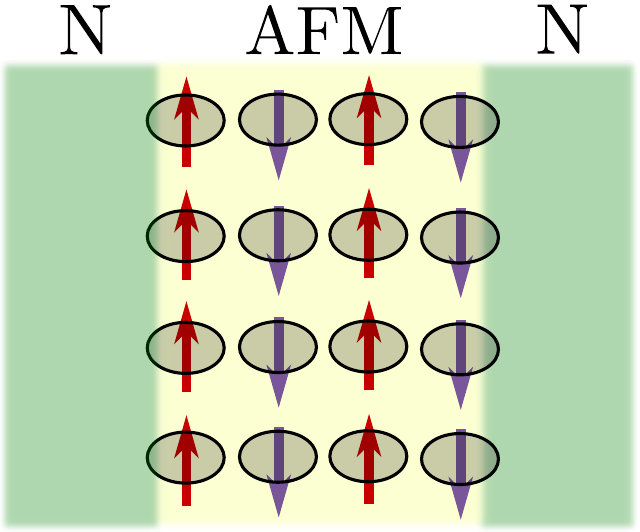}
		\caption{Schematic depiction of a trilayer heterostructure that allows coupling the two antiferromagnetic sublattices to two different normal metals.}
		\label{fig:uncomp_double}
	\end{center}
\end{figure}

The von Neumann entropy is widely employed as a measure to quantify entanglement between two subsystems. Thus, its value depends on how a larger system is partitioned into its entangled constituents. In the case of quantum spin liquids, it is common to draw an imaginary boundary and partition the magnet spatially into an inside and outside regions. The entanglement entropy may then be evaluated between these two spatial regions and allows to determine the entangled and/or topological nature of the ground state as well as excitations. On the other hand, in the case of two-mode squeezed states, the participating modes provide a natural partitioning for entanglement~\cite{Gerry2004}. The participating modes are entangled, which may be exploited for useful protocols~\cite{Gerry2004}. However, to this end, it is crucial to access the two entangled modes separately. 

As discussed in the main text, antiferromagnetic squeezed-magnons are comprised by the two-mode squeezing of the sublattice magnons. Therefore, in order to utilize the squeezing-mediated intrinsic entanglement between the sublattice-magnons, it is important to access the sublattice magnons individually. This can be achieved by employing AFMs with two uncompensated interfaces in a trilayer structure as depicted in Fig. \ref{fig:uncomp_double}. Similar heterostructures have also been proposed to host magnon-mediated indirect exciton condensation~\cite{Johansen2019}. The experimental methods and relevant materials for achieving uncompensated interfaces have been discussed elsewhere~\cite{Kamra2018A}. Furthermore, the recently discovered layered van der Waals AFMs~\cite{MacNeill2019} provide another promising route towards achieving the desired couping to the two sublattices. While Fig. \ref{fig:uncomp_double} depicts the example of coupling two normal metals to the antiferromagnetic sublattices, the general objective is to couple the two systems to be entangled, that are not necessarily metals, to the opposite sublattices.

\bibliography{SqAFM}

\end{document}